\documentclass[journal]{IEEEtran}
\usepackage[cmex10]{amsmath}
\usepackage{amssymb}
\usepackage{cite}
\usepackage{graphicx}
\usepackage{array,color}
\usepackage{multirow}
\usepackage{amsmath}
\usepackage{stfloats}
\usepackage{graphicx}
\usepackage{subfigure}
\usepackage{tabularx}
\usepackage{epsfig,epsf,color,balance,cite}
\usepackage{setspace}
\usepackage{bm}
\usepackage{textcomp}
\usepackage{algorithmic}
\usepackage{algorithm}
\usepackage{comment}
\usepackage{subfigure}
\usepackage{caption}
\usepackage{graphicx}
\usepackage{setspace}
\usepackage{diagbox}

\usepackage{multirow}
\usepackage{graphicx}
\makeatletter

\newcommand{\Rmnum}[1]{\expandafter\@slowromancap\romannumeral #1@}
\makeatother
\usepackage{booktabs}
\usepackage{amsmath}

\makeatother
\pdfoutput=1
\begin{document}

\title{Deep Learning Based Beam Training for Extremely Large-Scale Massive MIMO in Near-Field Domain}

\author{Wang Liu, Hong Ren, Cunhua Pan, and Jiangzhou Wang, $\textit{Fellow IEEE}$
	\thanks{ W. Liu, H. Ren and C. Pan are with National Mobile Communications Research Laboratory, Southeast University, Nanjing, China. (e-mail:{wangliu, hren, cpan}@seu.edu.cn). J. Wang is with the School of Engineering, University of Kent, UK. (e-mail:j.z.wang@kent.ac.uk). \emph{Corresponding author: Hong Ren, Cunhua Pan.}}
	\thanks{This work was supported in part by the National Natural Science Foundation of China (62101128) and Basic Research Project of Jiangsu Provincial Department of Science and Technology (BK20210205).}
}

\maketitle

\begin{abstract}
Extremely large-scale massive multiple-input-multiple-output (XL-MIMO) is regarded as a promising technology for next-generation communication systems. In order to enhance the beamforming gains, codebook-based beam training is widely adopted in XL-MIMO systems. However, in XL-MIMO systems, the near-field domain expands, and near-field codebook should be adopted for beam training, which significantly increases the pilot overhead. To tackle this problem, we propose a deep learning-based beam training scheme where the near-field channel model and the near-field codebook are considered. To be specific, we first utilize the received signals corresponding to the far-field wide beams  to estimate the optimal near-field beam. Two training schemes are proposed, namely the proposed original and the improved neural networks. The original scheme estimates the optimal near-field codeword directly based on the output of the neural networks. By contrast, the improved scheme performs additional beam testing, which can significantly improve the performance of beam training. Finally, the simulation results show that our proposed schemes can significantly reduce the training overhead in the near-field domain and achieve beamforming gains.

\end{abstract}


\vspace{-0.4cm}
\section{Introduction}

Compared to massive multiple input multiple output (MIMO), extremely large-scale massive MIMO (XL-MIMO) can further improve spectral efficiency for achieving higher beamforming gain through directional beamforming to compensate for severe path loss \cite{XLMIMO1}. Thus, XL-MIMO is a promising technology for the future sixth generation system. Furthermore,  millimetre-wave (mmWave) communications allow massive antenna arrays to be deployed at base stations (BSs), which guarantees the feasibility of XL-MIMO.

For mmWave massive MIMO systems, the beam training method based on predefined codebooks has been widely adopted. However, this will incur extremely high training overhead\cite{beam_sweep}. To tackle this issue, a typical scheme based on hierarchical codebooks that contain wide beams and narrow beams was proposed in \cite{h_codebok_1,h_codebok_2,h_codebok_3}.
Specifically, the hierarchical codebooks based scheme firstly searches for the optimal wide beam and then searches for the optimal narrow beam covered by the wide beam.

Owing to the enhanced feature of solving non-linear problems by deep learning, various deep learning-based beam training schemes have been proposed in \cite{make1,make2, qi}. Qi et al. proposed a deep neural network (DNN) based scheme where the received signals corresponding to a few beams are taken as the input for DNN to estimate the beam that best matches the channel\cite{qi}. Based on the received signals corresponding to wide beams, a long short-term memory (LSTM) network was proposed for mobile scenarios\cite{make1}, where previous received signals are also fed into the LSTM to enhance the accuracy of beam training. By contrast, \cite{make2} employed the previous state information of the low-frequency channel as the input for the LSTM network to predict the optimal beam in the high-frequency band.

For XL-MIMO systems, the extremely large-scale antenna arrays increase the Rayleigh distance, which incurs the near-field effect\cite{r_distance}. However, the above contributions are based on the far-field domain. In the near-field domain, codebook design needs to consider both the angle and the distance from the BS to the user equipment (UE), which is different from the far-field domain. Recently, the codebook design scheme based on the near-field domain was proposed in \cite{Daill_2}. This scheme increases the elements of the sampling distance, and thus the number of candidate beams also increases significantly. Therefore, it is crucial to reduce the overhead of codebook-based beam training in the near-field domain.

To tackle the above issue, we propose a deep learning-based beam training scheme for the near-field domain in XL-MIMO systems. Based on the received signals of the far-field wide beam, we design two neural networks to estimate the angle and distance of the optimal near-field beam, respectively. Then, the original and improved schemes are proposed. The former estimates the optimal near-field codeword directly based on the output of the neural networks. By contrast, the latter makes full use of the output of the neural networks by performing additional beam test, which can significantly improve the performance of beam training. 

The paper is organized as follows. The near-field channel model and beam training model are presented in Section \ref{systemmodel1}. Section \ref{DL_model} presents the deep learning model and the proposed two beam training schemes. Section \ref{simulation} provides the simulation results. Conclusions are drawn in  Section \ref{conclusion}.

\vspace{-0.2cm}
\section{System Model}\label{systemmodel1}
\subsection{Channel Model}
Consider an uplink mmWave XL-MIMO single-user communication system in the near-field environment. The BS is equipped with $N$ antennas and the UE has a single antenna. Both the BS and the UE are assumed to have only a single radio frequency (RF) chain, which means that only analog combining is used during the communication. For simplicity, uniform linear arrays (ULAs) are employed at the BS. The near-field channel model can be represented as\cite{Daill_2}
\vspace{-0.1cm}
\begin{eqnarray}\label{nf_ch_model}
\boldsymbol{h}= \sqrt{{N}/{L}}\sum\nolimits_{l=1}^{L}g_{l}e^{-j\frac{2\pi }{\lambda _{c}}r_{l}}\mathbf{b}\left ( \theta _{l},r_{l} \right ),
\end{eqnarray}
where $\lambda_{c}$ and $L$ denote the wavelength and the number of channel paths, respectively;   $g_{l}$, $r_{l}$, and $\theta_{l}$  denote the channel gain, the distance, and the angle of the $l$-th path, respectively. Specifically, $\theta_{l}$ is given by $\theta_{l} = \frac{2d}{\lambda _{c}}\textrm{sin}\Theta _{l}$, where $\Theta_{l}$ denotes the physcial angle-of-arrival (AOA) and $d$ denotes the antenna spacing at the BS. For simplicity, we set $d=\lambda /2$. Furthermore, the near-field steering vector $\mathbf{b}\in\mathbb{C}^{N\times 1}$ can be expressed as
\vspace{-0.2cm}
\begin{eqnarray}\label{nf_str_vector}
\mathbf{b}\!\left(\theta_{l}, r_{l}\right)\!=\!\!\frac{1}{\sqrt{N}}\!\!\left[e^{-j \frac{2\pi }{\lambda _{c}}\left(r_{l}^{(0)}\!-\!r_{l}\right)}, \cdots, e^{-j \frac{2\pi }{\lambda _{c}}\left(r_{l}^{(N-1)}\!-\!r_{l}\right)}\!\!\right]^{T}\!\!,
\end{eqnarray}
where $r_{l}^{(n)}$ denotes the distance between the $n$-th BS antenna and the UE or scatter.

\vspace{-0.3cm}
\subsection{Beam Training Model}

We assume that a predefined codebook is applied for beamforming. As illustrated in (\ref{nf_str_vector}), the angular-domain discrete Fourier transform (DFT) codebook for the far-field domain will no longer be applicable to the near field. In order to increase the distance sampling of near-field paths in the codebook, \cite{Daill_2} proposed a polar coordinate domain-based codebook as
\vspace{-0.1cm}
\begin{align}\label{nf_codebook}
{\cal W} = &\left\{ {{\bf{b}}\left( {{\theta _1},r_1^1} \right), \ldots ,{\bf{b}}\left( {{\theta _N},r_1^N} \right),  } \right.\nonumber\\
&\;\;\;\;{\rm{                  }}\ldots\left. {{\bf{b}}\left( {{\theta _1},r_S^1} \right), \ldots ,{\bf{b}}\left( {{\theta _N},r_S^N} \right)} \right\},
\end{align}
where each element is a candidate beam codeword, $\theta_{n} $ denotes the $n$-th sampled angle, and  $r_{s}^{n}$ denotes the $n$-th sampled distance on the $s$-th distance ring.

\begin{figure}
	\begin{minipage}[t]{0.49\linewidth}
		\centering
		\includegraphics[width=1.4in]{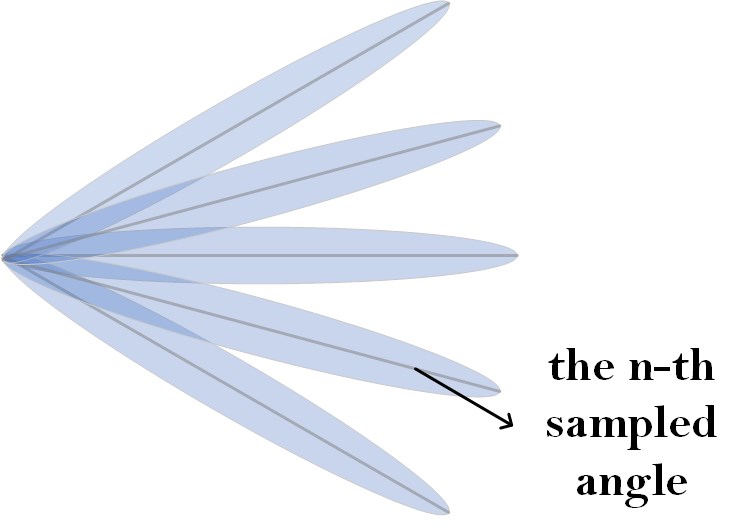}
		\vspace{-0.2cm}
		\caption{\small The far-field angular-domain codebook.} 
		\label{maben2}\vspace{-0.7cm}
	\end{minipage}%
	\hfill
	\begin{minipage}[t]{0.49\linewidth}
		\centering
		\includegraphics[width=1.4in]{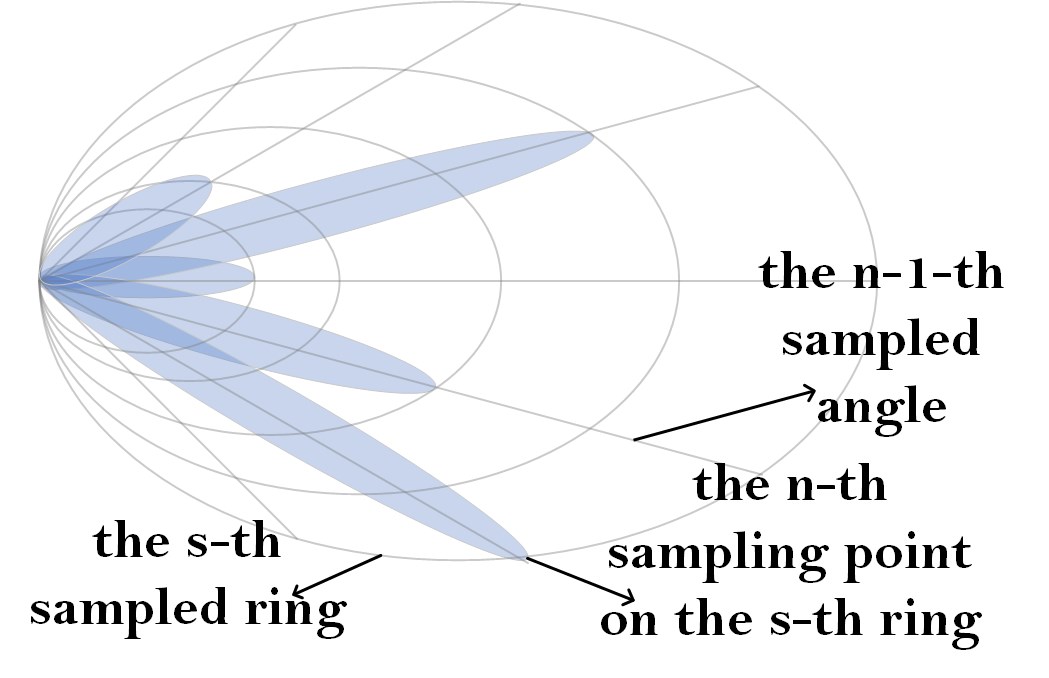}
		\vspace{-0.2cm}
		\caption{\small The near-field polar-domain codebook.} 
		\label{maben1}\vspace{-0.7cm}
	\end{minipage}%
	\hfill
\end{figure}

Similar to the far-field codebook in Fig. \ref{maben2}, the near-field codebook employs uniform sampling for angular sampling with $N$ equal parts as shown in Fig. \ref{maben1}. Different from the far-field codebook, the near-field codebook also performs the distance sampling and the whole space is divided into $S$ distance rings. For the near-field case, the distance from the sampling point to the BS is  $r_{s}^{n}$, where $s=1,2,\cdots,S$, $n=1,2,\cdots,N$. Thus, the near-field codebook contains $I= NS$ codewords for this scheme and the $i$-th codeword is defined as
\begin{equation}\label{codeword}
\begin{gathered}
\boldsymbol{w}_{i}=\mathbf{b}\left(\theta_{n}, r_{s}^{n}\right), i=\left ( s-1 \right )N+n.
\end{gathered}
\end{equation}

Then, the received signal at the BS can be written as
\begin{eqnarray}\label{re_signal}
y=\sqrt{P}\boldsymbol{w}^{H}\boldsymbol{h}x+\boldsymbol{w}^{H}\boldsymbol{n},
\end{eqnarray}
where $P$, $\boldsymbol{w}\in \mathcal{W}$ and $\boldsymbol{n}\sim \mathcal{C N}\left(\mathbf{0}, \sigma^{2} \boldsymbol{I}_{N}\right)$ denote the transmit power,  the analog combining beam and the additive white Gaussian noise (AWGN) vector, respectively. Assuming that the data signal symbols satisfy $\left|x \right|^{2}=1$, the achievable rate can be expressed as
\begin{eqnarray}\label{re_rate}
R=\log _{2}\left(1+{P\left|\boldsymbol{w}^{H} \boldsymbol{h} \right|^{2}}/{\sigma^{2}}\right).
\end{eqnarray}

Due to the severe path loss of mmWave communications,  the received signal power mainly comes from the line-of-sight (LoS) path. Therefore, our aim is to identify the optimal codeword from the predefined codebook to align the beam to the LoS path and thus achieve the largest data rate. Therefore, the codeword selection problem can be formulated as
\vspace{-0.2cm}
\begin{eqnarray}\label{problem}
	\boldsymbol{w}^{\ast }=\mathop{\arg\max}\limits_{\boldsymbol{w} \in \mathcal{W}}\;\;  \log _{2}\left(1+{P\left|\boldsymbol{w}^{H} \boldsymbol{h} \right|^{2}}/{\sigma^{2}}\right).
\end{eqnarray}

A straightforward solution is beam sweeping, which can obtain the highest achievable rate. However, it will also incur a huge training overhead compared with that used in the far-field scenario. In the far-field domain, a hierarchical codebook-based beam training scheme was developed in  \cite{h_codebok_1}. Wide beam codewords and narrow beam codewords are generally included in hierarchical codebooks, and the beams formed are referred to as wide beams and narrow beams, respectively. The $n$-th narrow beam codeword can be expressed as
\vspace{-0.2cm}
\begin{eqnarray}\label{narr_beam}
\boldsymbol{w}_{n}^{\textrm{n}}=\sqrt{{1}/{N}}\left[1, e^{\mathrm{j} \pi \sin \theta _{n}^{\textrm{n}}  }, \cdots ,e^{\mathrm{j} \pi\left(N-1\right) \sin \theta _{n}^{\textrm{n}} }\right]^{\mathrm{T}},
\end{eqnarray}
where $\theta _{n}^{\textrm{n}}$ denotes the beam angle of the $n$-th narrow beam. The $m$-th wide beam codeword can be expressed as
\vspace{-0.1cm}

\begin{eqnarray}\label{wide_beam}
	\boldsymbol{w}_{m}^{\textrm{w}}=\sqrt{\frac{1}{N/T}}\left[1, e^{\mathrm{j} \pi \sin\theta _{m}^{\textrm{w}}} ,\cdots, e^{\mathrm{j} \pi\left(\frac{N}{T}-1\right) \sin \theta _{m}^{\textrm{w}} }\right]^{\mathrm{T}},
\end{eqnarray}
where $T$ is the number of narrow beams each large beam contains and $\theta _{m}^{\textrm{w}}$ is the angle of the $m$-th wide beam. In addition, to generate a wide beam, some antennas need to be deactivated at the BS. Specifically, only $N/T$ antennas are employed to generate the wide beam. Correspondingly, the number of wide beam codewords is only $M=N/T$.
	
The wide beam formed by a wide beam codeword can cover $T$ narrow beams formed by the narrow beam codeword as a high-level beam. As a result, a single wide beam can cover a larger range of angles, and the number of wide beam codewords is typically reduced. In the hierarchical beam training scheme, all wide beam codewords will be tested in the first phase to identify the best wide beam, and the narrow beam covered by that wide beam will be tested in the second phase to obtain the optimal narrow beam codeword. Compared to the beam sweeping technique, the hierarchical codebook scheme dramatically reduces the training and pilot overhead.

\section{Deep Learning for Beam Training}\label{DL_model}
\subsection{Problem Formulation}

 Intuitively, we can utilize a portion of beams to find the optimal beam codeword with low training overhead. However, it will cause degraded performance when the direction of the LoS path is different from the selected narrow beams. Motivated by the hierarchical codebook, we propose to measure the received signals of far-field wide beams to find the optimal codeword in the near-field codebook. The received signals corresponding to far-field wide beams provide information on the angle and distance of the LoS path, which can be utilized to find the optimal near-field beam. As the relationship between the received signal and the parameters of the LoS path is highly nonlinear, we turn to use the deep learning method for beam training. Specifically, we train the direction neural network and distance neural network to estimate the angle index $n$ and the ring index $s$ of the optimal near-field codeword based on the received signal corresponding to the far-field wide beams, respectively. The received signal of the $m$-th far-field wide beam is given by
\vspace{-0.1cm}
\begin{eqnarray}\label{singal_w}
y_{m}^{\textrm{w}}=\sqrt{P}\boldsymbol{w}_{m}^{\textrm{w}H}\boldsymbol{h}x+\boldsymbol{w}_{m}^{\textrm{w}H}\boldsymbol{n}.
\end{eqnarray}

Thus, the singal vector consisting of all received singals corresponding to the far-field wide beams can be expressed as
\begin{eqnarray}\label{singal_w_set1}
\mathbf{y}^{\textrm{w}}=\left [ y_{1}^{\textrm{w}},y_{2}^{\textrm{w}},\cdots ,y_{M}^{\textrm{w}} \right ]^{T}.
\end{eqnarray}

Therefore, the deep learning based near-field beam training problem can be formulated as
\vspace{-0.1cm}
\begin{equation}\label{dl_model}
	\begin{gathered}
	n^{\ast }=\mathit{f}_{1}\left ( \mathbf{y}^{\textrm{w}} \right ),n^{\ast }\in \left\{1,2,\cdots ,N \right\},\\
	s^{\ast }=\mathit{f}_{2}\left ( \mathbf{y}^{\textrm{w}} \right ),s^{\ast }\in \left\{1,2,\cdots ,S \right\},
	\end{gathered}
\end{equation}
where $\mathit{f}_{1}$ and $\mathit{f}_{2}$ denote the mapping functions of direction network and distance network, respectively; $n^{\ast }$, $s^{\ast }$ denote the optimal angle index and the optimal ring index in the near-field codebook. According to (\ref{dl_model}), the index of the optimal near-field codeword can be formulated as
\vspace{-0.1cm}
\begin{eqnarray}\label{singal_w_set}
\vspace{-0.1cm}
	i^{\ast }=\left ( s^{\ast }-1 \right )N+n^{\ast }.
\end{eqnarray}

The proposed deep learning scheme has training and estimation stages. During the training stage, we generate a large number of data sets which are used to train the neural network's parameters. The BS receives pilot signals using all wide beams and feeds the received signals into a well-trained neural network model, which predicts the index of the optimal codeword at the estimation stage.

\vspace{-0.5cm}
\subsection{Deep Learning Model Design}

\begin{figure}
	\centering
	\includegraphics[width=3.1in]{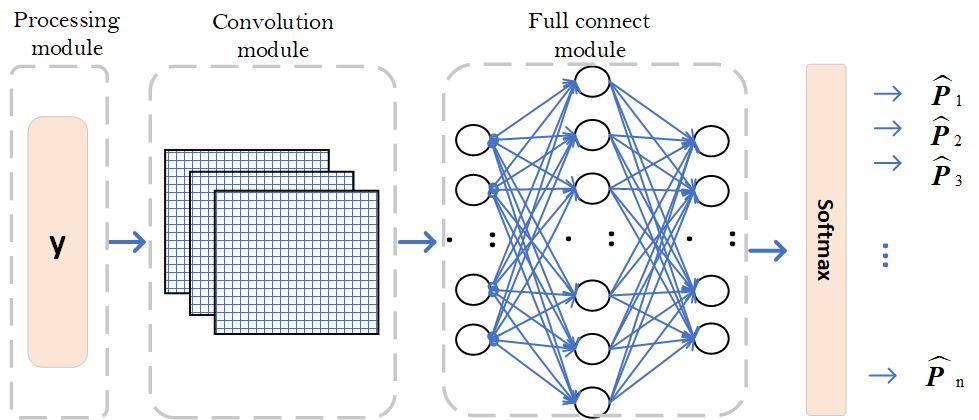}
	\vspace{-0.3cm}
	\caption{\small Proposed neural networks structure for beam training . }
	\label{fig1}\vspace{-0.8cm}
\end{figure}

As shown in Fig. \ref{fig1}, the proposed neural network model consists of four modules: the input module, the convolution module, the fully connected module, and the output module. The convolutional module contains convolutional neural networks, i. e., convolutional neural networks (CNN), which has powerful feature extraction capabilities\cite{CNN_ability}. It is worth noting that the direction and distance networks have the same input, convolution and fully connected module as in Fig. \ref{fig1}, and they only differ in the output module because the numbers of direction and distance samplings are different. The reason for considering two different neural networks is that estimating the distance index and the direction index requires the CNN to extract different features from the received signal.


$1)$ $\textit{Input Modoule}$: According to  (\ref{dl_model}), the signal vector $\mathbf{y}^{\textrm{w}}$ is used as input to the neural network. 
 Firstly, the signal is divided into real part $\mathfrak{R}\left ( \mathbf{y}^{\textrm{w}} \right )$ and imaginary part $\mathfrak{J}\left ( \mathbf{y}^{\textrm{w}} \right )$, and then the real and imaginary parts are transformed into a matrix form, which can be fed to the convolution module.

$2)$ $\textit{ Convolution Module}$: Multiple convolutional layers are included in the convolution module, each of which is followed by a ReLu activation layer and a normalization layer. The ReLu activation layer's goal is to improve the neural network's ability to handle nonlinear mappings, while the normalization layer's goal is to speed up the neural network's convergence and avoid overfitting. 

$3)$ $\textit{  Fully Connected Module}$: Multiple hidden layers are embedded in the fully connected module. Similar to convolutional layers, each hidden layer is followed by a ReLu activation layer and a normalization layer. The function of the fully-connected layers is to refine the features extracted from the convolution module.

$4)$ $\textit{ Output Module}$:
The Softmax activation layers are utilized in the output module to transform the raw output of the neural network into a probability distribution vector.
Specifically, the outputs of the direction network and the distance network can be written as
\begin{eqnarray}\label{out}
\hat{\boldsymbol{P}}^{\textrm{a}}=\left [ \hat{p}_{1}^{\textrm{a}},\hat{p}_{2}^{\textrm{a}},\cdots ,\hat{p}_{N}^{\textrm{a}}\right ]^{T},
\hat{\boldsymbol{P}}^{\textrm{r}}=\left [ \hat{p}_{1}^{\textrm{r}},\hat{p}_{2}^{\textrm{r}},\cdots ,\hat{p}_{S}^{\textrm{r}}\right ]^{T},
\end{eqnarray}
where $\hat{p}_{n}^{\textrm{a}}$ and $\hat{p}_{s}^{\textrm{r}}$ denote the estimated probability of the optimal near-field codeword on the $n$-th angle and the $s$-th ditance ring. The larger the value of $\hat{p}_{n}^{\textrm{a}}$ and $\hat{p}_{s}^{\textrm{r}}$, the higher the probability that the optimal codeword is on the $n$-th angle and on the $s$-th distance ring.

Since the role of the proposed neural network is to find the best angle index and ring index, its function is the same as a classifier. Therefore, the cross-entropy loss function is applied to train the proposed neural network. The cross-entropy loss functions of direction and distance networks can be written as
\vspace{-0.1cm}
\begin{equation}\label{loss} {\text{Loss}}^{\textrm{a}}\!\!=\!\!-\!\!\sum\nolimits_{n=1}^{N}\!\!p_{n}^{\textrm{a}}\log_{10}\hat{p}_{n}^{\textrm{a}},{\text{Loss}}^{\textrm{r}}\!\!=\!\!-\!\!\sum\nolimits_{s=1}^{S}p_{s}^{\textrm{r}}\log_{10}\hat{p}_{s}^{\textrm{r}},
\end{equation}
where $p_{n}^{\textrm{a}}=1$ and ${p}_{s}^{\textrm{r}}=1$ if the actual optimal near-field codeword is on the $n$-th angle and the $s$-th distance ring. Otherwise, $p_{n}^{\textrm{a}}=0$ and ${p}_{s}^{\textrm{r}}=0$.


\vspace{-0.3cm}
\subsection{Deep Learning Based Beam Training }
In this section, we develop two beam training schemes. 

$1)$ $\textit{ Original Scheme}$: We first use all far-field wide  beams to receive the pilot signals  sent by the UE. By testing all the far-field wide beams, we can obtain the received signal vector $\mathbf{y}^{\textrm{w}}$ according to (\ref{singal_w}) and (\ref{singal_w_set}). Then, we feed the received signal vector $\mathbf{y}^{\textrm{w}}$ to direction network and distance network to obtain  probability distribution vector $\hat{\boldsymbol{P}}^{\textrm{a}}$ and $\hat{\boldsymbol{P}}^{\textrm{r}}$ accroding to (\ref{out}). Based on $\hat{\boldsymbol{P}}^{\textrm{a}}$ and $\hat{\boldsymbol{P}}^{\textrm{r}}$ , the angle index and distance ring index of the optimal near-field codeword can be estimated as
\begin{equation}\label{index_or1}
	\begin{gathered}
		n^{\ast }=\arg\max_{n=1,2,\cdots ,N}[\hat{\boldsymbol{P}}^{\textrm{a}}]_{n},
		s^{\ast }=\arg\max_{s=1,2,\cdots ,S}[\hat{\boldsymbol{P}}^{\textrm{r}}]_{s}.
	\end{gathered}
\end{equation}
The index of the optimal codeword can be obtained as
\begin{eqnarray}\label{index_or21}
	i^{\ast }=\left ( s^{\ast }-1 \right )N+n^{\ast }.
\end{eqnarray}
Then, the obtained optimal near-field codeword  is
\begin{eqnarray}\label{index_or22}
\vspace{-0.2cm}
	\boldsymbol{w}^{\ast }=[\mathcal{W}]_{i^{\ast }}.
\end{eqnarray}

\vspace{-0.5cm}
\begin{algorithm}[H]
    \setstretch{0.7}
	\caption{The Original Scheme}\label{alg:alg1}
	\begin{algorithmic}[1]
        \STATE {\textbf{Input:}} well-trained direction and distance networks.

		\STATE \hspace{0.5cm}Perform wide beam training tests: receive pilot signals  with all wide beams and obtain $\mathbf{y}^{\textrm{w}}$.
		\STATE \hspace{0.5cm}Feed $\mathbf{y}^{\textrm{w}}$ into the distance and direction network, obtain $\hat{\boldsymbol{P}}^{\textrm{a}}$ and $\hat{\boldsymbol{P}}^{\textrm{r}}$ .
		\STATE \hspace{0.5cm}Obtain the optimal angle index $n^{\ast }$ and ring index $s^{\ast }$ via (\ref{index_or1}).
		\STATE \hspace{0.5cm}Obtain the index of the optimal near-field codeword $	i^{\ast }$ in the codebook via (\ref{index_or21}).
		\STATE \hspace{0.5cm}Obtain the optimal near-field codeword $\boldsymbol{w}^{\ast }$ via (\ref{index_or22}).
		\STATE {\textbf{Output:}} $\boldsymbol{w}^{\ast }$
	\end{algorithmic}
\end{algorithm}
\vspace{-0.4cm}

$2)$ $\textit{ Improved Scheme}$: Based on the same steps as in the original scheme, the received signal vector $\mathbf{y}^{\textrm{w}}$  and the probability distribution vectors $\hat{\boldsymbol{P}}^{\textrm{a}}$ and $\hat{\boldsymbol{P}}^{\textrm{r}}$  are obtained in the improved scheme, where the BS still employs far-field wide beams to receive signals. After obtaining $\hat{\boldsymbol{P}}^{\textrm{a}}$ and $\hat{\boldsymbol{P}}^{\textrm{r}}$, the improved scheme performs additional tests on a small number of near-field codewords with the largest several estimated probabilities, during which the UE  needs to send additional pilot signals. Additional test will further enhance the efficiency of beam training. Based on $\hat{\boldsymbol{P}}^{\textrm{a}}$ and $\hat{\boldsymbol{P}}^{\textrm{r}}$, we can obtain the $K$ most possible angle indices and the $L$ most possible ring indices that should be clearly emerged in $\mathcal{L}_{\mathrm{a}}$ and $	\mathcal{L}_{\mathrm{d}}$, respectively. Mathematically, it can be formulated as
\vspace{-0.2cm}
\begin{align}\label{order}
	&\left\{ \hat{p}_{\sigma _{1}}^{\textrm{a}},\hat{p}_{\sigma_{2}}^{\textrm{a}},\cdots ,\hat{p}_{\sigma _{N}}^{\textrm{a}}\right\}=\left< \left\{\hat{p}_{1}^{\textrm{a}},\hat{p}_{2}^{\textrm{a}},\cdots ,\hat{p}_{N}^{\textrm{a}} \right\}\right>,
		\\
	&\left\{ \hat{p}_{\gamma _{1}}^{\textrm{r}},\hat{p}_{\gamma_{2}}^{\textrm{r}},\cdots ,\hat{p}_{\gamma _{S}}^{\textrm{r}}\right\}=\left< \left\{\hat{p}_{1}^{\textrm{r}},\hat{p}_{2}^{\textrm{r}},\cdots ,\hat{p}_{S}^{\textrm{r}} \right\}\right>,
	\\
	&\mathcal{L}_{\mathrm{a}}=\left \{ \sigma _{1},\sigma _{2},\cdots ,\sigma _{K} \right \},
	\mathcal{L}_{\mathrm{d}}=\left \{ \gamma _{1},\gamma _{2},\cdots ,\gamma _{L} \right \},
\end{align}
where $\left \langle \cdot  \right \rangle$ denotes the  order operation, e.g., for $\mathcal{A}=\left\{a_{1}, a_{2}, \ldots, a_{n}\right\},\langle\mathcal{A}\rangle=\left\{a_{\sigma_{1}}, a_{\sigma_{2}}, \ldots, a_{\sigma_{n}}\right\}$ with $a_{\sigma_{1}} \geq a_{\sigma_{2}} \geq \ldots \geq a_{\sigma_{n}}$. Then, the set of indices at the intersection of these angles and rings can be expressed as
\begin{eqnarray}\label{index_or3}
	\mathcal{B} = \left\{{b}_{j}\big|{b}_{j}=(\gamma -1)N+\sigma ,\sigma \in 	\mathcal{L}_{\mathrm{a}},\gamma \in \mathcal{L}_{\mathrm{d}}\right\}.
\end{eqnarray}
The set of near-field codewords that requires additional test is
\begin{eqnarray}\label{index_or4}
\mathcal{T} = \left \{ \boldsymbol{w}_{b_{1}},\boldsymbol{w}_{b_{2}},\cdots ,\boldsymbol{w}_{b_{KL}} \right \}.
\end{eqnarray}
By implementing additional near-field beam tests, where the near-field beam comes from the set $\mathcal{T}$, we can obtain the additional received singal vector
\begin{align}\label{ad_re_signal}
	\mathbf{y}^{\textrm{T}}&=\left [y_{b_{1}},y_{b_{2}} ,\cdots , y_{b_{KL}}\right ]^{T},
\end{align}
where $y_{b_{j}}\!\!\!=\!\!\!\sqrt{P}\boldsymbol{w}_{b_{j}}^{H}\boldsymbol{h}x\!+\!\boldsymbol{w}_{b_{j}}^{H}\boldsymbol{n}$. Finally, the index of the optimal near-field codeword in codebook $\mathcal{W}$ can be formulated as
\begin{eqnarray}\label{index_or51}
\vspace{-0.1cm}
i^{\ast }=\mathop{\arg\max}\limits_{b_{j}\in \mathcal{B}}\left | y_{b_{j}} \right |.
\end{eqnarray}
Then, the optimal near-field codeword can be expressed as
\begin{eqnarray}\label{index_or52}
\vspace{-0.1cm}
\boldsymbol{w}^{\ast }=[\mathcal{W}]_{i^{\ast }}.
\end{eqnarray}

Note that when the user can only see a portion of the base station antennas due to the excessive number of MIMO antennas, we can reduce the dimensionality of the codebook to accommodate the reduction in the number of available antennas\cite{review4}. 

\begin{algorithm}
    \setstretch{0.75}
	\caption{ The Improved Scheme}\label{alg:alg2}
	\begin{algorithmic}[1]
		\STATE {\textbf{Input:}} well-trained direction and distance network, $K$, $L$
		
		\STATE \hspace{0.5cm}Perform wide beam training tests: receive pilot  with all wide beams and obtain $\mathbf{y}^{\textrm{w}}$.
		\STATE \hspace{0.5cm}Feed $\mathbf{y}^{\textrm{w}}$ into the distance and direction network, obtain $\hat{\boldsymbol{P}}^{\textrm{a}}$ and $\hat{\boldsymbol{P}}^{\textrm{r}}$ .
		\STATE \hspace{0.5cm}Obtain the $K$ most probable angle indice $\mathcal{L}_{\mathrm{a}}$ and the $L$ most probable ring indice $\mathcal{L}_{\mathrm{d}}$  via (\ref{order}).
		\STATE \hspace{0.5cm}Obtain the set of indice at the intersection $\mathcal{B}$ via (\ref{index_or3}), and the near-field codewords that require additional testing $\mathcal{T}$ via (\ref{index_or4}).
		\STATE \hspace{0.5cm}Perform $KL$ additional near-field beam traing tests: receive pilot with near-field beams in $\mathcal{T}$ and  obtain $\mathbf{y}^{\mathcal{T}}$ via (\ref{ad_re_signal}).
		\STATE \hspace{0.5cm}Obtain the index of the optimal near-field codeword $	i^{\ast }$ in codebook via (\ref{index_or51}).
		\STATE \hspace{0.5cm}Obtain the optimal near-field codeword $\boldsymbol{w}^{\ast }$ via (\ref{index_or52}).
		\STATE {\textbf{Output:}} $\boldsymbol{w}^{\ast }$
	\end{algorithmic}
\end{algorithm}

\vspace{-0.6cm}


\vspace{-0.2cm}
\section{Simulation Results}\label{simulation}
 We consider an mmWave  XL-MIMO communication system. The number of channel paths $L$ is set to 3, which includes one LoS path with channel gain $g_{1} \sim \mathcal{C} \mathcal{N}(0,1)$ and two NLoS paths with channel gains $g_{2}, g_{3} \sim \mathcal{C} \mathcal{N}(0,0.01)$.  In addition, the number of antennas of the BS is equal to 512. The distance from the UE to the BS  is selected from $\mathcal{U}(10 \mathrm{~m}, 60 \mathrm{~m})$, which satisfies the near-field condition. Using such settings and codebook design scheme in \cite{Daill_2}, the number of sampling angles is $N=512$, and the number of sampling distance rings is $S=5$. Thus, the number of codewords in the near-field codebook is $I=2560$.

For the neural network model, the detailed network structure and parameters can be shown in Table \ref{table2}. $C_{\text{in}}$ and $C_{\text{out}}$ of the convolution module denote the input and output feature channels, respectively. $C_{\text{in}}$ and $C_{\text{out}}$ of the fully connected module denote the dimension of the input vector and the output vector for that layer, respectively.The kernel sizes in convolution layers are denoted by $(w_{1},w_{2},w_{3})$. For the offline training phase, we generate 100,000 sets of labelled data, of which $80\%$ are used to train the neural network, and 10\% for the validation and test sets. Furthermore, the test set is separated from the validation and training sets for testing the accuracy of the deep neural network. The batch size and the initial learning rate are set to 1000 and 0.01, respectively.

Two metrics are adopted to evaluate the performance: 1) Normalizd signal to noise ratio (SNR): 
$	G_{N}\!\!\!\triangleq\!\!\!{\left|\hat{\boldsymbol{w}}^{\ast H} \boldsymbol{h} \right|^{2}}/{\left|\boldsymbol{w}^{\ast H} \boldsymbol{h} \right|^{2}}$, where $\boldsymbol{w}^{\ast H}$ and $\hat{\boldsymbol{w}}^{\ast H}$ are the actual and estimated optimal near-field codeword, respectively.
2) Effective achieveable rate
$\bar{E}\triangleq\left ( 1\!\!-\!\! {T_{\mathrm{tra}}}/{T_{\text {tot }}} \right )\log _{2}\left(1\!\!+\!\!{P\left|\hat{\boldsymbol{w}}^{\ast H}  \boldsymbol{h} \right|^{2}}/{\sigma^{2}}\right)$,
where $T_{\mathrm{tot}}$ is the channel coherence time and  $T_{\text {tra }}$ is equal to the product of the number of tested beams and test time $t_s$ per beam \cite{ts_time}.

\begin{table}[h]
    \footnotesize
    \setstretch{1.1}
	\centering 
	\caption{\text { \footnotesize NETWORK PARAMETERS }} 
    \vspace{-0.3cm}
	\label{table2} 
	\begin{tabular}{|c|c|c|}
		\hline
		\textbf{Moudle}                  & \textbf{Layer}  & \textbf{Parameter}                       \\ \hline
		\multirow{5}{*}{\!\!Convolution\!\!}  & Convolution  & $C_{\text {in}}$=2, $C_{\text {out}}$=64, (3,3,3)    \\ \cline{2-3}
		& \!\!ReLu activation\!\!\! & $\!\!C_{\text {in}}$=64, $C_{\text {out}}$=64\!\!\!\!                \\ \cline{2-3}
		& \!\!Convolution\!\!\!     & $\!\!C_{\text {in}}$=64, $C_{\text {in}}$=256, (3,1,1)\!\!\!\!       \\ \cline{2-3}
		& \!\!ReLu activation\!\!\! & $\!\!C_{\text {in}}$=64, $C_{\text {out}}$=256\!\!\!\!               \\ \cline{2-3}
		& \!\!Pooling         & $\!\!C_{\text {in}}$=256, $C_{\text {out}}$=256\!\!\!\! \\ \hline
		\multirow{6}{*}{\!\!\!Fully Connected\!\!\!} & Hidden          & $C_{\text{in}}$=256, $C_{\text{out}}$=1024     \\ \cline{2-3}
		& \!\!ReLu activation\!\!\! & $\!\!C_{\text {in}}$=1024, $C_{\text {out}}$=1024\!\!\!\!            \\ \cline{2-3}
		& \!\!Hidden\!\!\!          & $\!\!C_{\text {in}}$=1024, $C_{\text {out}}$=1024\!\!\!\!             \\ \cline{2-3}
		& \!\!ReLu activation\!\!\! & $\!\!C_{\text {in}}$=1024, $C_{\text {out}}$=1024\!\!\!\!           \\ \cline{2-3}
		& \!\!Hidden\!\!\!          & $\!\!C_{\text {in}}$=1024, $C_{\text {out}}$=512\!\!\!\!              \\ \cline{2-3}
		& \!\!ReLu activation\!\!\! & $\!\!C_{\text {in}}$=512, $C_{\text {out}}$=512\!\!\!\!              \\ \hline
	\end{tabular}
\vspace{-0.3cm}
\end{table}

\begin{figure}[h]
\vspace{-0.4cm}
		\centering
		\includegraphics[width=2.4in]{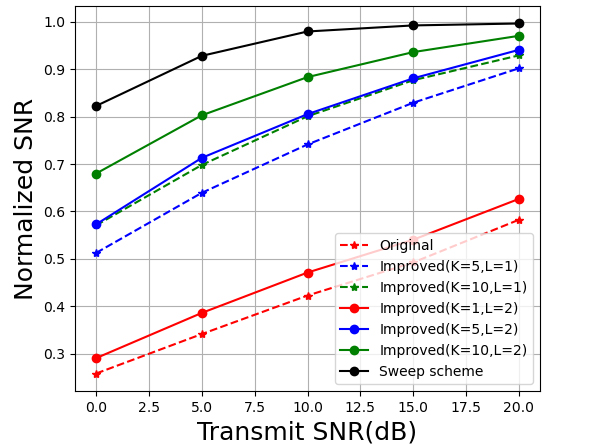}
       \vspace{-0.2cm}
		\caption{\small Normalized SNR for the original and improved schemes. }
		\label{fig3}\vspace{-0.3cm}
\end{figure}

Fig. \ref{fig3} shows the performance of two proposed schemes in terms of normalized SNR
$G_{N}$. It can be seen that the improved scheme always achieves better performance in terms of the achievable rates as it can make full use of the probability distribution vector of the neural network output. As expected, the SNR of the improved scheme increases with $K$ and $L$. When  the SNR is larger than 10 dB for the case of $K$=10 and $L$=2, the improved scheme can achieve over 90\% of the data rate using the perfect beam training. However, the improved scheme only requires $M+KL=148$ beams, much less than the naive sweeping scheme that needs 2560 beams. As for the improved scheme, the number of additional beams that need to be tested is usually less than 20. As a result, the two proposed schemes can reduce the pilot overhead by approximately 95\%.

\begin{figure}[h]
		\centering
		\includegraphics[width=2.4in]{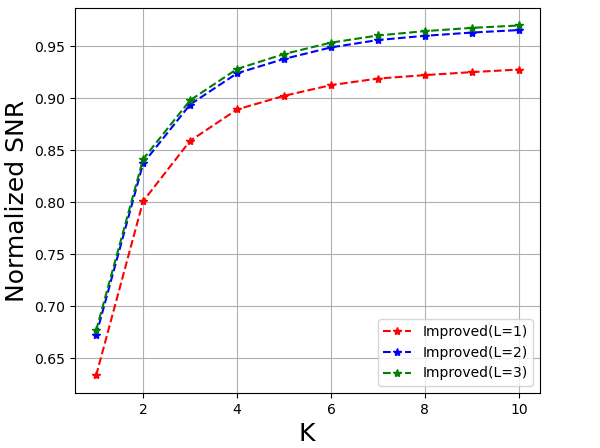}
        \vspace{-0.3cm}
		\caption{\small Normalized SNR for the improved scheme.}
		\label{fig4}\vspace{-0.3cm}
\end{figure}

Fig. \ref{fig4} illustrates the impact of different values of $K$ and $L$ on the performance of the improved scheme. When $K$ is less than 5, the normalized SNR $G_{N}$ increases rapidly with $K$. However, when $K$ is larger than 5,  $G_{N}$  tends to converge. In addition, the normalized SNR $G_{N}$ also increases with $L$.
\vspace{-0.4cm}

\begin{figure}[h]
       \vspace{-0.3cm}
	\centering
	\includegraphics[width=2.4in]{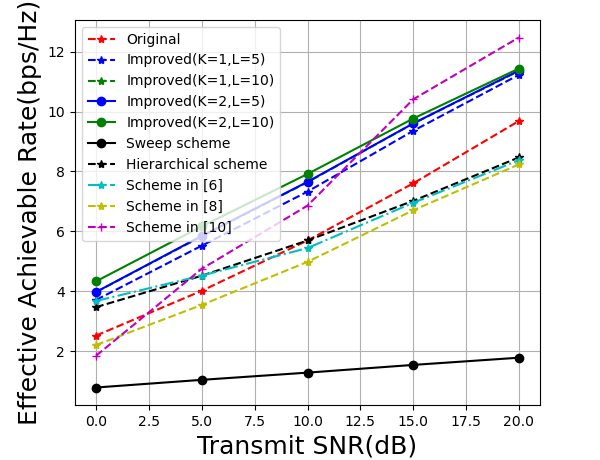}
       \vspace{-0.2cm}
	\caption{\small Effective achievable rate v.s. Transmit SNR (dB)}
	\label{fig6}\vspace{-0.7cm}
\end{figure}

Fig. \ref{fig6} compares the two proposed schemes with the existing beam training and channel estimation schemes in terms of effective achievable rate $\bar{E}$. From Fig. \ref{fig6}, we can see that the proposed scheme can achieve better performance than the far-field beam training schemes, which means that the beam selected by our proposed beam training schemes can match the near-field channel significantly better. Furthermore, the improved scheme is also superior to the near-field channel estimation scheme proposed in [10] under the condition of low SNR and achieve similar performance at high SNR. It can also be seen from Fig. \ref{fig6} that although the improved scheme requires additional beam tests compared to the original scheme, higher effective achievable rates are obtained.


\section{Conclusion}\label{conclusion}
We proposed a deep learning-based beam training approach to reduce the training overhead for  XL-MIMO systems, where the near-field domain was considered. Two neural networks were designed to estimate the angle and distance. Then, we developed original and improved schemes. The original scheme estimates the optimal near-field codeword directly based on the output of the neural networks, whilst the improved scheme performs additional beam testings. Simulation results showed that our proposed schemes can significantly reduce the training overhead.

\vspace{-0.3cm}
\bibliographystyle{IEEEtran}
\bibliography{myre}


\end{document}